AN INAUGURAL LECTURE TO THE SOCIETY

# EARTHQUAKE-INDUCED LANDSLIDE HAZARDS IN MOUNTAIN REGIONS: A REVIEW OF CASE HISTORIES FROM CENTRAL ASIA

Hans-Balder HAVENITH & Céline BOURDEAU

(11 figures, 1 table)

*Géorisques et Environnement, Département de Géologie, Université de Liège,
B-4000 Sart Tilman - Liège. E-mail : HB.Havenith@ulg.ac.be*

**ABSTRACT** This paper presents a summary of the main trigger factors of earthquake-induced landslides as well as a review of case histories of major landslide-triggering earthquake events in Central Asia. The Kainama earth-flow case history of 2005 is added to document possible mid-term effects of smaller earthquakes. These events show that in the Central Asian Mountains, two types of seismically triggered mass movements may have particularly disastrous effects: massive long runout rockslides and medium-sized earth flows made of loess – or a mixture of both. These types of mass movements also significantly contributed to the largest natural catastrophe of the last century in Central Asian mountain regions: the 1949 Khait earthquake.
The high impact potential of these types of mass movements is further pointed out through comparison with two worldwide known events, the 1920 Haiyuan (China) and the 1976 Peru earthquake.
Case studies had been carried out on rockslides, debris slumps and earth flows triggered by the above-mentioned Kemin and Suusamyr earthquakes as well as other smaller seismic shocks in the Kyrgyz Tien Shan. Many of the investigated landslides had known a complex failure history before final collapse.
To better assess the short- to long-term effects of earthquakes on slopes, landslides need to be surveyed more intensively, over mid- and long-terms.

**RESUME** Aléas de glissements de terrain induits par tremblements de terre dans les régions montagneuses: Exemples d'Asie Centrale. Nous présentons, dans cet article, les facteurs principaux qui contribuent au déclenchement d'instabilités sismiques à travers des exemples de ruptures majeures en Asie Centrale. S'ajoute à cela la description d'un glissement de terrain (Kainama) induit par des séismes de plus faible magnitude. Ces événements montrent que dans les montagnes d'Asie Centrale, essentiellement deux types de mouvements de masse d'origine sismique peuvent avoir un impact dévastateur : il s'agit des grands 'glissements rocheux' et des 'coulées de loess' (ou éventuellement une combinaison de ces deux types). Ces types de mouvements de masse ont également causé les plus grands impacts de l'événement le plus catastrophique dans les montagnes d'Asie Centrale du siècle passé: le tremblement de terre de Khait de 1949.
Dans cet article, l'impact potentiel des instabilités d'origine sismique en Asie Centrale est analysé au moyen d'une comparaison avec des mouvements de masse induits par deux événements majeurs : le séisme de Chine de 1920 et le séisme du Pérou en 1976.
Nous avons analysé en détail les glissements rocheux, les glissements de débris et les écoulements de terre engendrés par les séismes de Kemin, de Suusamyr et des séismes de plus faible magnitude dans le Tien Shan kirghize. Parmi ces mouvements de masse, beaucoup résultent d'une longue et complexe histoire de rupture.
Nous pensons qu'une instrumentation plus approfondie de quelques instabilités existantes permettrait de mieux comprendre les effets à moyen et à long terme des séismes sur la stabilité des versants.

**MOTS-CLES** : Mouvements de masse, facteurs sismo-tectoniques et géologiques, effets de site, liquéfaction, effets à long-terme, montagnes du Tien Shan et du Pamir

## 1. Introduction

During the last ten years, after a series of disastrous earthquake events in mountain regions in Taiwan (1999), El Salvador (2001), Pakistan (2005) and China (2008), increasing attention has been addressed to landslides triggered by earthquakes.

Previously, landslides have been considered as minor effects of earthquakes compared to the impact of the ground shaking itself. Schuster & Lynn (2001) partly attributed the perception of the relatively small impact of earthquake-triggered mass movements to the fact that many related losses are often referred to as direct consequences of the earthquake. The assumption of the global 'secondary' importance of earthquake-induced landslides and other ground failures was confirmed by a study undertaken by Bird&Bommer (2004) – even though they also admit that, for some earthquake risk scenarios, seismic ground failures need to be taken into consideration to correctly estimate total losses. Explicitly, it was only noticed for the two El Salvador earthquakes in 2001 that 'the most devastating impact … has been the triggering of hundreds of landslides in volcanic soils, …' (Bommer et al., 2002).

This study of Bird & Bommer (2004) had been completed well before the M=7.6 earthquake hit the Kashmir mountains on October 8, 2005. For this event, Petley et al. (2006) estimated that about 30% of the total number of killed people (officially 87350), i.e. 26500, had been victims of co-seismic landslides. Less than three years later, on May 12, 2008, the Wenchuan earthquake hit the Sichuan and neighbouring provinces of China and caused 'more than 15000 geohazards in the form of landslides, rockfalls, and debris flows, which resulted in about 20000 deaths' (Yin et al. 2009). These deaths represent again almost 30 % of the official total number of fatalities of 69197 of this event.

In this paper, we will evaluate the potential impact of landside hazards induced by earthquakes on the basis of a review of related geological observations. However, we do not aim at statistically reviewing earthquake-induced landslides events, related trigger mechanisms and impacts. For this, we refer to the works done by Keefer (1984 ad 1999) and Rodriguez et al. (1999).

Here, focus will be on the relatively poorly known Central Asian mountain regions, the Tien Shan and Pamir, on the basis of landslide case histories related to the following earthquakes (see location in Fig. 1 and a summary in Table 1): M=8.2 Kemin, 1911, M=7.6 Sarez, 1911, M=7.4 Khait, 1949, M=5.5 Gissar, 1989 and M=7.3 Suusamyr, 1992. One landslide case history is added to document the mid-term effect of small and medium-sized earthquakes on slope stability and related hazards in the Tien Shan Mountains.

Two other M>7 earthquakes of the last century in the Tien Shan Mountains will not be further documented here since we know only little: the M=7-7.4 Chatkal earthquake in Western Tien Shan, for which some data about triggered mass movements and landslide dams are provided by Leonov (1965), and the M=7.3 Markansu earthquake affecting remote areas of the Southern Tien Shan, for which Nikonov et al (1983) indicated very few data about induced ground failures.

The documented earthquake case histories of Central Asia will be compared with some of the worldwide most disastrous events, and particularly with the M=8.5 Haiyuan (or Gansu, 1920) and the M=7.8 Peru (1970) earthquakes. These two events caused the greatest number of deaths in history through multiple triggered landslides and a single mass movement, respectively (Keefer, 1999).

Further, our analysis will focus on those types of mass movements, which caused these disasters: rapid flows

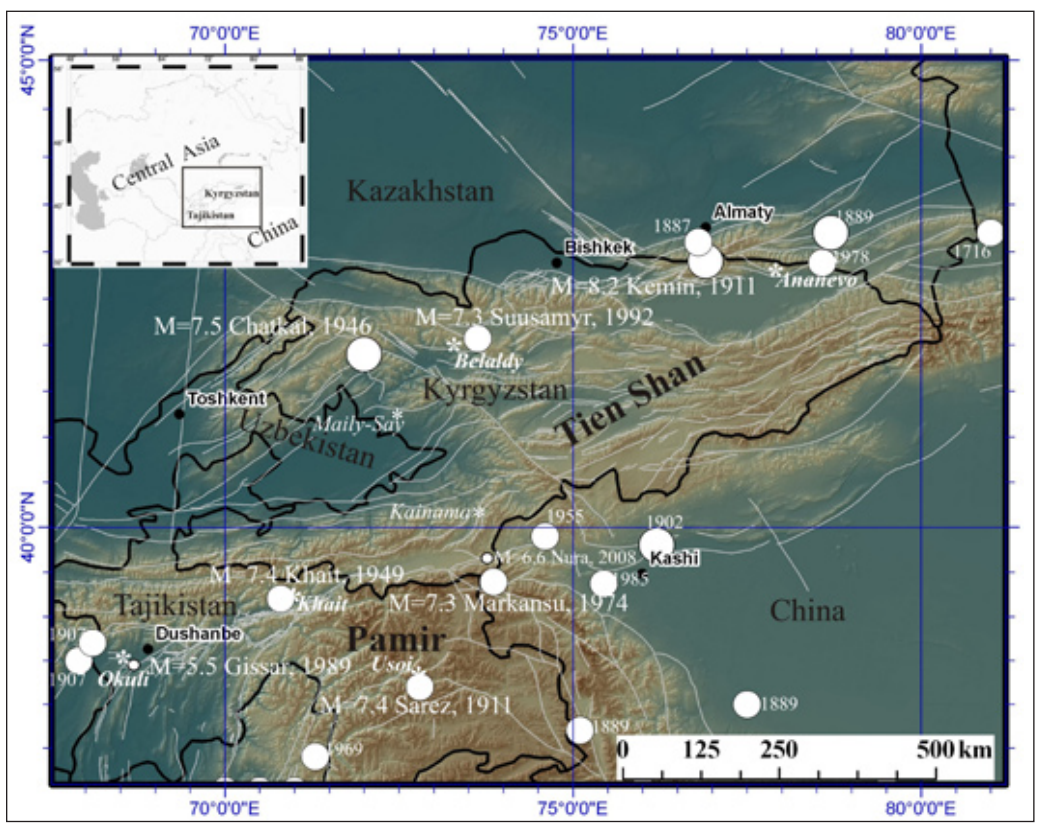

**Figure 1**: Map of Tien Shan and Pamir Mountains in Central Asia (general map in upper left corner) with location of major faults and earthquakes (white filled circles show all recorded M>=7 earthquakes with the year of occurrence; the magnitude is indicated for analysed events); earthquake-triggered mass movements and investigated landslide sites are marked by stars with names in italic.

| Earthquake | Year | Rockslides | Landslides in soft sediments/ loess |
|---|---|---|---|
| M=8.2 Kemin, Kyrgyzstan | 1911 | Chon-Kemin and Ananevo > $10 \cdot 10^6$ m$^3$ | Several small landslides |
| M=7.6 Sarez, Tajikistan | 1911 | Usoi ~ $2 \cdot 10^9$ m$^3$ | Not reported |
| M=7.4 Khait, Tajikistan | 1949 | Khait ~ $75 \cdot 10^6$ m$^3$ | Yasman loess earth-flow |
| M=5.5 Gisar, Tajikistan | 1989 | none | Okuli loess earth-flow and other loess slides |
| M=7.3 Suusamyr, Kyrgyzstan | 1992 | Belaldy ~ $40 \cdot 10^6$ m$^3$ | Chet-Korumdy debris-slides |

Table 1. Investigated landslide-triggering earthquakes in the Tien Shan and Pamir.

(mainly) in loess deposits and massive rock avalanches. Actually, those types also induce the highest geological risk in Central Asian regions, such as shown in the following. For a complete review of landslide types triggered by earthquakes – including submarine landslides -, the reader is referred to Keefer (1984, 1999) and Rodriguez et al. (1999).

Special attention will be paid to long-term effects of earthquakes in mountain regions. Examples will be shown for clearly delayed triggering of slope failures after earthquakes and post-seismic increase of landslide activity, such as observed after the Chi-Chi earthquake of 1999 in Taiwan (Dadson et al., 2004). In this regard, we will also analyse previously called 'secondary or tertiary effects' of earthquakes, such as natural dams and related flooding impacts. In the Tien Shan, the most recent massive natural dam was formed after the Suusamyr earthquake in 1992 – it partly failed in 1993, thus causing a long-runout debris flow and widespread flooding downstream. The 2008 Wenchuan earthquake clearly marked the importance of such effects – which could have killed another few thousands of people if efficient mitigation measures had not been taken by Chinese authorities. For these cases, it will be shown that the term 'consequential hazard' proposed by Korup (2003) should be preferred over 'secondary of tertiary effects' implying a lesser degree of importance.

## 2. Geological and morphological setting and earthquake-induced landslide types

This section analyses *where* landslides are preferentially triggered by earthquakes within seismically active mountain belts with respect to seismo-tectonic, geological-structural and morphological factors.

For a classification of seismically triggered landslide types we refer to the extensive work of Keefer (1984, 1999). Here, we will just introduce the general categories of earthquake-triggered landslides defined by Keefer (1999). These include 1) 'Disrupted Slides and Falls', 2) 'Coherent Slides and Landslides', 3) 'Lateral Spreads and Flows'. As mentioned above, for Central Asian mountain regions, massive rockslides and rock avalanches as well as loess earth-flows are the most hazardous types of mass movements. These belong, respectively, to the first and third category of rapid mass movements.

### 2.1. Seismo-Tectonic factors

According to Keefer (1984) and Rodríguez et al. (1999), landslides may be triggered by earthquakes with a magnitude equal or larger than 4.

Sassa (1996) further observed that seismic landslide occurrence is strongly dependent on the proximity of the fault rupture. More recently, Yin et al. (2009) highlighted that 'the distribution of landslide-dammed lakes is closely related to proximity to the earthquake fault in the Longmenshan Mountain system. There were 22 landslide-dammed lakes distributed within the earthquake fault belt, which is 1/3 of the total number of landslide-dammed lakes'. All over China, Wen et al. (2004) observed that 'the locations of most of the giant landslides are very close to the major fault zones… particularly the active earthquake zones'. After the Chi-Chi earthquake (1999) in Taiwan, Khazai & Sitar (2003) outlined that 'the majority of the landslides occurred within a distance of 20 km from the fault rupture plane on the hanging wall on south and southeast facing slopes. Less than 10% of the landslides occurred on the footwall.' Similarly, for the Kashmir earthquake in 2005, Petley et al. (2006) noted that 'close to the fault rupture there was a high incidence of landslides triggered by the Kashmir earthquake... The distribution of landslides appears to be very asymmetric, with most of the landslides being located on the hanging wall (north-eastern) side of the fault'. For the same event, Sato et al. (2007) further noticed that more than one third of the landslides occurred within 1 km from the active fault. Korjenkov et al. (2004) studying ground failures in the Suusamyr region, Northern Tien Shan, observed that 'seismic deformations are concentrated to active Neotectonic border faults, especially to their hanging limbs …and steep limbs of anticline-like uplifts'.

### 2.2 Geological-structural factors

The geological factors have also been analysed by Keefer (1984, 1999) who concluded that any geological material with low geomechanical strength, be it soil or rock, may be susceptible to earthquake-induced slope instability. The low strength may be related to weak cementation, intense weathering or fracturing, high water saturation or poor compaction. Sensitive clays, sand and loess are particularly susceptible to seismic ground failure.

With respect to the influence of the rock/sediment structure on seismic slope stability, Khazai & Sitar (2003) observed for landslides triggered by the 1999 Chi-Chi earthquake that 'in contrast to the shallow slides that typically occurred along joints and fractures oblique to the bedding/ foliation, the deep-seated failures appeared to be mostly sub parallel to bedding and/or foliation…'

For giant landslides in China, Wen et al. (2004) noted that 'rainfall-induced landslides more often occurred in soils with slip surfaces along the soil–rock contacts, and in the stratified rocks with contrast in competency in which slip surfaces occurred along bedding planes, while earthquake-induced landslides were more prevalent in stratified competent rocks in which slip surfaces occurred along cross-bedding joints…'

Most soft sediment landslides triggered by earthquakes in the Tien Shan seem to have failed along bedding contacts (Havenith et al., 2003a).

From these observations, no clear trend of local structural effects can be outlined – especially for seismic rock slope failure, it is not clear if cross-bedding joints or bedding contacts are more prone to the development of a sliding surface. Here, clearly more research based on structural geology mapping is needed.

## 2.3. Morphological factor

Earthquake-induced landslides may be triggered from any surface morphology – even within flat areas, such as lateral spreads, or from steep cliffs, such as rock falls.

Still, several particularities can be outlined. From observations after the 2005 Pakistan earthquake, Sato et al. (2007) inferred that 'there was a slight trend that large landslides occurred on vertically convex slopes rather than on concave slopes; furthermore, large landslides occurred on steeper (30° and more) slopes than on gentler slopes'. This topographic effect was also revealed by Harp et al. (2006) noting that '… several ridges near the top also are covered with fractures that are concentrated at the ridge crests and the summit'. They conclude that 'these fractures probably are associated with increased levels of shaking due to topographic amplification of the ground shaking'.

Zhang & Wang (2007) noted that 'landslides induced by rainfall occurred mainly on steep slopes, whereas those induced by the Haiyuan Earthquake mainly occurred on relatively gentler slopes, indicative that the mechanisms of these two types of landslides are very different.'

From the above citations and our own experience of earthquake-induced landslides in the Tien Shan (see below), it can be concluded that mainly the surface curvature has an influence on seismic slope stability at a global scale; particularly, hillcrests, higher parts of slopes and convex surface morphologies are prone to seismic slope failure. The influence of slope angle on seismic slope stability is not clear; in some cases, especially in rocks, steeper slopes are more prone to instability; in others, especially in soft sediments, gentle slopes produce most of the mass movements, indicating that the combined effect of slope and geology has to be taken into consideration. Similarly the influence of the slope aspect on slope stability is closely connected to local environmental and tectonic conditions.

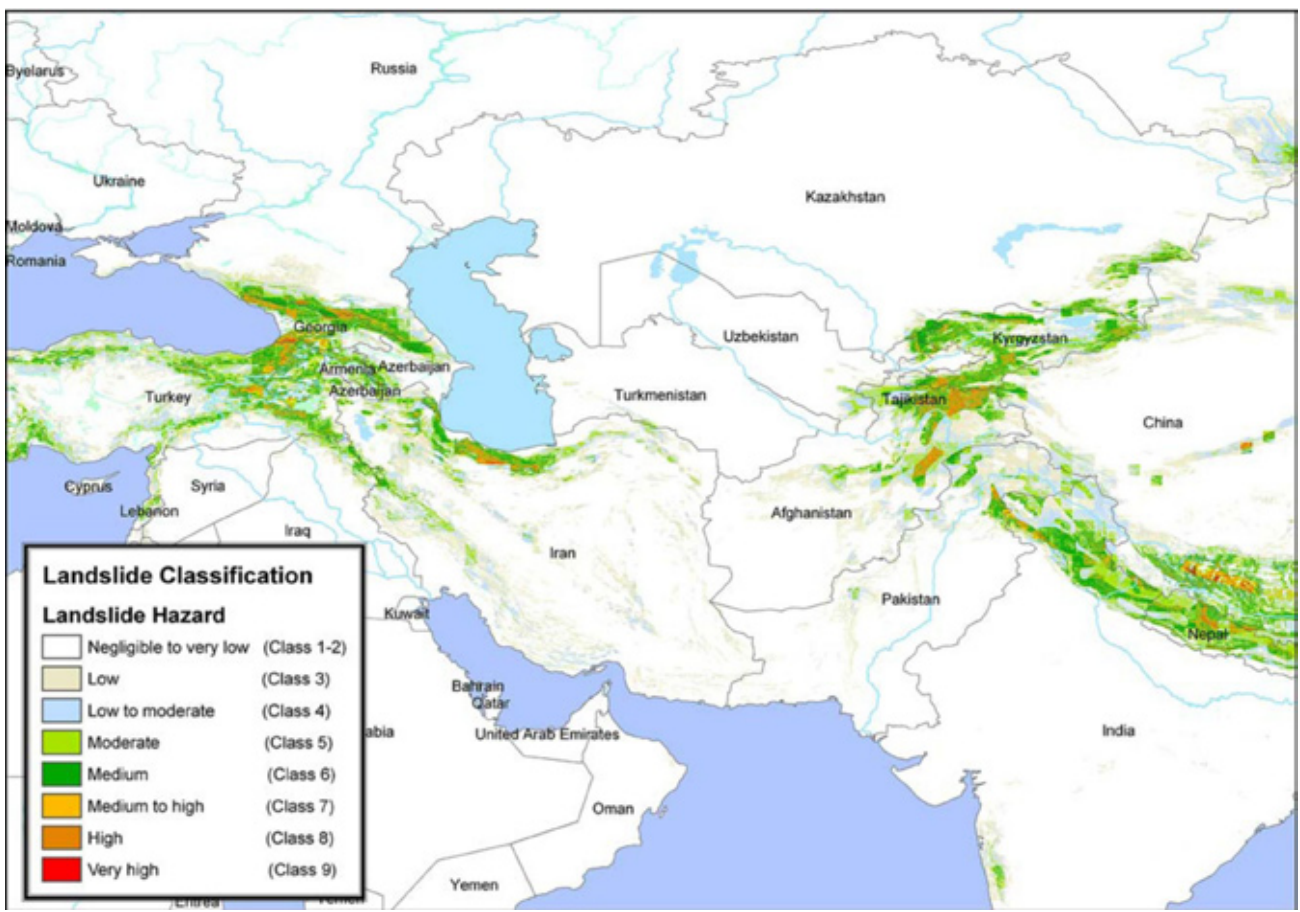

**Figure 2**: Landslide hazard map of Central Asia and Middle East (from Nadim et al., 2006).

## 3. Case histories from Central Asia compared with events in China and Peru

Nadim et al. (2006) assessed landslide and avalanche occurrence probabilities worldwide on the basis of morphological, geological, meteorological and seismological data. They clearly showed that all landslide hotspots are located in seismically active mountain ranges. For Central Asia (Fig. 2), they estimate that global landslide hazard can be rated as medium to very high. They further noted that some areas in Tajikistan are marked by highest mortality risk due to landslides.

The following case histories document the landslide hazard and risk triggered by earthquakes in the Tien Shan and the Pamir Mountains. A comparison will be made with the 1920 Haiyuan (China) and 1970 Peru events to outline the most important factors contributing to landslide hazard and risk in Central Asia.

### 3.1. Case histories from Central Asia

#### 3.1.1. The Kemin earthquake, 1911

The Kemin Ms = 8.2 earthquake of 1911 (January 3) is one of the strongest events ever recorded in the Tien Shan; it was first analysed by Bogdanovich et al. (1914). The earthquake caused extensive landsliding along the activated fault segments over a length of 200 km. The largest mass movements were two rockslides, one within the Kemin valley and the other north of the lake Issyk-Kul. The first rock avalanche (about 15 $10^6$ m3) made of limestone material occurred along the activated Chon Kemin fault at about 60 km W of the epicentre, and is known to have buried a village of yourts with 38 inhabitants. The second 'Ananevo' rockslide located in the north of lake Issyk Kul (at some 80 km east of the presumed epicentre) is one of the most prominent features produced by the Kemin earthquake (Havenith et al., 2002, see Fig. 3). Failure took place at the southern end of a mountain ridge, just above the discontinuous Chon Aksu fault also activated by the 1911 Kemin earthquake. According to Delvaux et al. (2001), this section of the Chon Aksu fault is a thrust gently dipping towards the northeast into the collapsed slope. Evidence of the presence of the fault is the related scarp with a height of 1 m at 3 km WNW of the site increasing up to almost 10 m at 12 km to the WNW. On the site itself, outcrops at the foot of the southwest-oriented slope show particularly disintegrated and weathered granitic rocks within a 100-200 m thick fault zone.

#### 3.1.2. The Sarez earthquake, 1911

The Sarez earthquake, Ms=7.6, struck the central Pamir Mountains, Tajikistan, on February 18, 1911. Such an earthquake is likely to have triggered hundreds or thousands of mass movements, but only one is well documented: the giant Usoi rockslide (Fig. 4), which fell from a 4500 m high mountain down to an elevation of 2,700 m in the valley (Schuster & Alford, 2004). This rockslide has formed a dam with a volume of about 2 $10^9$ m$^3$ on Murgab River. According to Schuster & Alford (2004), the location of the slide is related to 'a high degree of rock fracturing from previous tectonic activity…a major thrust fault with an unfavourable orientation … and… a series of intensively sheared zones forming geometric setting for a typical wedge failure.'

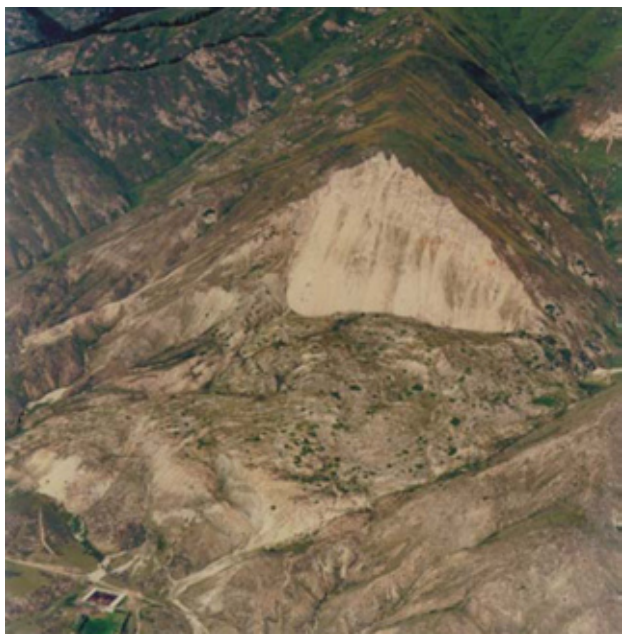

**Figure 3:** Helicopter Photograph of the Ananevo rockslide (10-15 $10^6$ m$^3$). Scale is given by the horse-farm building (lower left corner) of roughly 60 by 60 m.

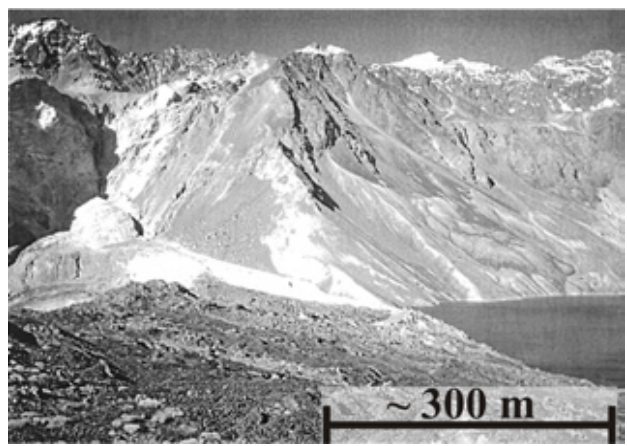

**Figure 4:** Usoi rockslide scarp (left) and dam (front) with dammed Sarez lake (right), after Schuster & Alford (2004).

Behind this 600 m high natural dam (the highest dam in the world), Lake Sarez and the smaller Lake Shadau had been impounded, the first one with a maximum depth of 500.

Various scenarios consider the risk related to the failure of the rockslide dam due to internal erosion or overtopping as well as to a flood wave induced by the impact of mass movement into Lake Sarez. Those scenarios are discussed by Schuster & Alford (2004) and Risley et al. (2006) – the worst-case scenario flood wave could affect more than 5 million people living in the Amu Darya river basin.

*3.1.3. The Khait earthquake, 1949*

According to Leonov (1960), Solonenko (1970) and Gubin (1960), the M=7.4 Khait earthquake that struck Northern Tajikistan on July 10, 1949, produced one of the most destructive earthquake-triggered landslide events in human history. Though, the western world knows relatively little about it – probably because most information has been published in Russian. First, a massive rock avalanche had buried the villages of Khait & Kusurak with up to one thousand inhabitants (Fig. 5); the exact number of fatalities will never be known since 'during the formidable rule of Joseph Stalin, information about accidents and natural catastrophes was suppressed unless special permission was granted' (Yablokov, 2001). This rock avalanche had been triggered from Borgulchak mountain at an altitude of about 2950 m and travelled more than 6 km before reaching the inhabited valley at an altitude of 1550 m. The volume was initially estimated to more than $200 \; 10^6 \; m^3$ (Leonov, 1960). However, more recent investigations by Evans et al. (2009) indicated that the total volume would be much lower, of about $75 \; 10^6 \; m^3$. Evans et al. (2009) also observed that a significant part of the mass movement was made of loess, which probably contributed to the mobility of the initial rockslide.

Second, Evans et al. (2009) indicate that in the Yasman valley opposite to the Khait rock avalanche, hundreds of loess earth-slides coalesced to form one massive earth-flow, which is believed to have buried about 20 villages. In total, the Khait rock avalanche as well as the Yasman loess earth-flows are likely to have killed more than 4000 people during the 1949 event.

In addition to the catastrophic impacts, Russian geologists described also the general conditions of the earthquake-triggered slope failures. For instance, Leonov (1960) wrote (translated from the Russian original): '… involved are also amplification effects that can explain landsliding far from the epicentre…' This observation had already been pointed out above.

*3.1.4. The Gissar earthquake, 1989*

South of Dushanbe, in Gissar, Tajikistan, a Ms=5.5 earthquake on January 23, 1989 had triggered a series of earth-flows in loess. At least 200 people were killed and hundreds of houses were buried. According to Ishihara et al. (1990), those slides were all related to extensive liquefaction, which had developed for a horizontal acceleration of about 0.15g. Ishihara et al. (1990) associated the liquefaction to the 'collapsible nature' of the highly porous loess material (a silt-sized deposit with an average content of clay of 15 % and a low plasticity).

The largest landslide, called 'Okuli' (Fig. 6), had an estimated volume of $20 \; 10^6 \; m^3$. Ishihara (2002) indicated that 'at least two slides seem to have been triggered independently from the hillsides on the north, which then merged into the main stream of the mudflow.' The sliding surface of most landslides was located at a depth of about 15 m within the saturated part of the 30 m thick loess deposits. Ishihara et al. (1990) also noted that the scarps of many landslides were located along a water channel installed on the shoulder of the hills. They assumed that 'water in the channel had been infiltrating in the loess over years leading to final failure during earthquake due to liquefaction of water-bearing loess layer'. This is supported by their observation of muddy water oozing from the earth-flow.

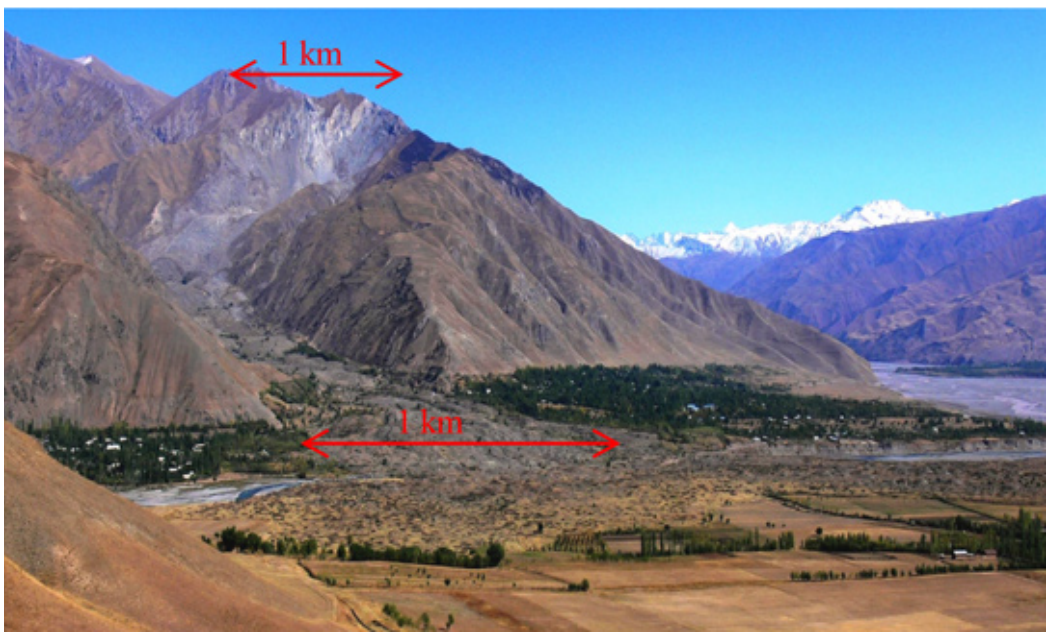

**Figure 5:** Khait rock avalanche; view towards the East from Yasman valley (colour version of photograph of 2005 provided by A. Ischuk, published in Evans et al., 2009).

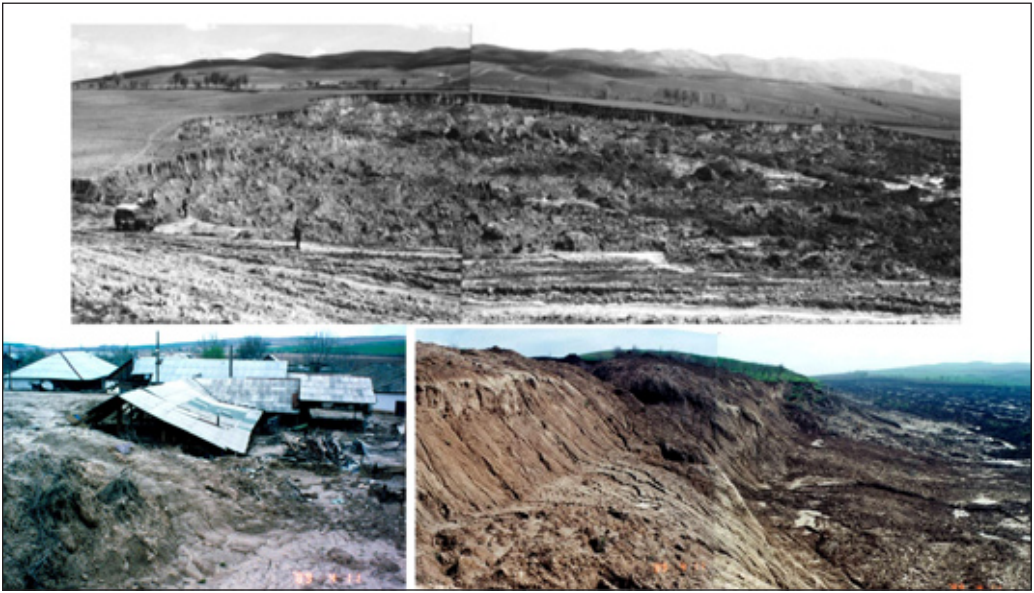

**Figure 6:** Okuli loess earth-flow. (photographs by Ishihara, 2002).

### 3.1.5. The Suusamyr earthquake, 1992

The most recent large seismic event hitting Central Asian mountain regions was the Ms=7.3 Suusamyr earthquake on August 19, 1992, triggering various types of ground failures in the Northern-Central Tien Shan (Bogachkin et al. 1997).

Most of the 50 people killed in the remote areas were victims of mass movements. Korjenkov et al. (2004) described a series of ground failures: sagging of mountain slopes, rockfalls, landslides, soil avalanches and flows, mud/debris flows, and also a great variety of gravitation cracks. Extensive ground failures could be observed along the crest and southern slope of the Chet-Korumdy ridge – here, most landslides had developed from previously existing ground instabilities (top left part of Fig. 7).

Most of the ground and slope failures spread over 4000 km² around the Suusamyr basin and the neighbouring mountain ranges were relatively small (less than $10^6$ m³). A map presenting all detected landslides in the Suusamyr region as well as ground failures induced by the Suusamyr earthquake is presented in Fig. 7 (modified from Korjenkov et al., 2004).

Only one large rockslide, the Belaldy rock avalanche, occurred on the southern slopes of the Suusamyr range (Bogachkin et al., 1997 and Korjenkov et al., 2004). It has covered a shepherd's family and a flock of sheep. The situation before and after the earthquake of the rockslide site is shown in Fig. 8. This mass movement had formed a dam on Jalpaksu River with a thickness of about 100 m, a width of 700 m, and volume of more than $40 \cdot 10^6$ m³. Behind the dam two small lakes were impounded (with an area of 200-300 m² in September 1992). In less than a year, the water level had increased enough to induce partial failure of the dam (Korjenkov et al., 2004). This failure resulted in a 20 km long mud- and debris-flow, which caused a lot of damage for infrastructure of Toktogul region. The space photograph of 1993 of the Belaldy site (Fig. 8) shows the upper part of the debris flow just below the dam.

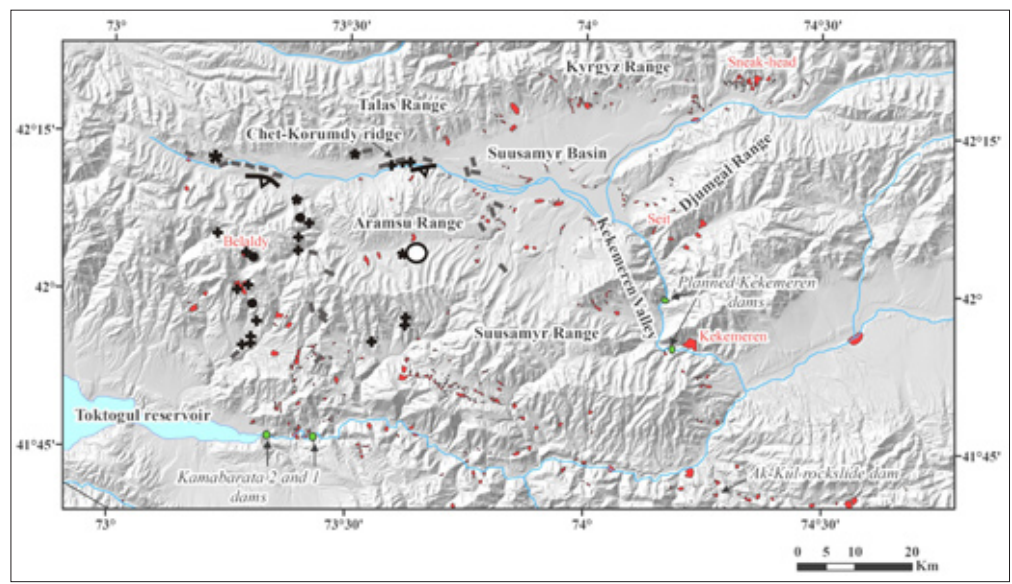

**Figure 7**: Map of all detected landslides in the Suusamyr region (red) and ground failures triggered by the Suusamyr earthquake (modified from Korjenkov et al., 2004 - circle: M=7.3 epicentre; black lines with triangle: 1992 fault scarps; crosses: rockslides and landslides; stars: mud eruptions; grey bars: ground and slope fractures, black dots: rock falls).

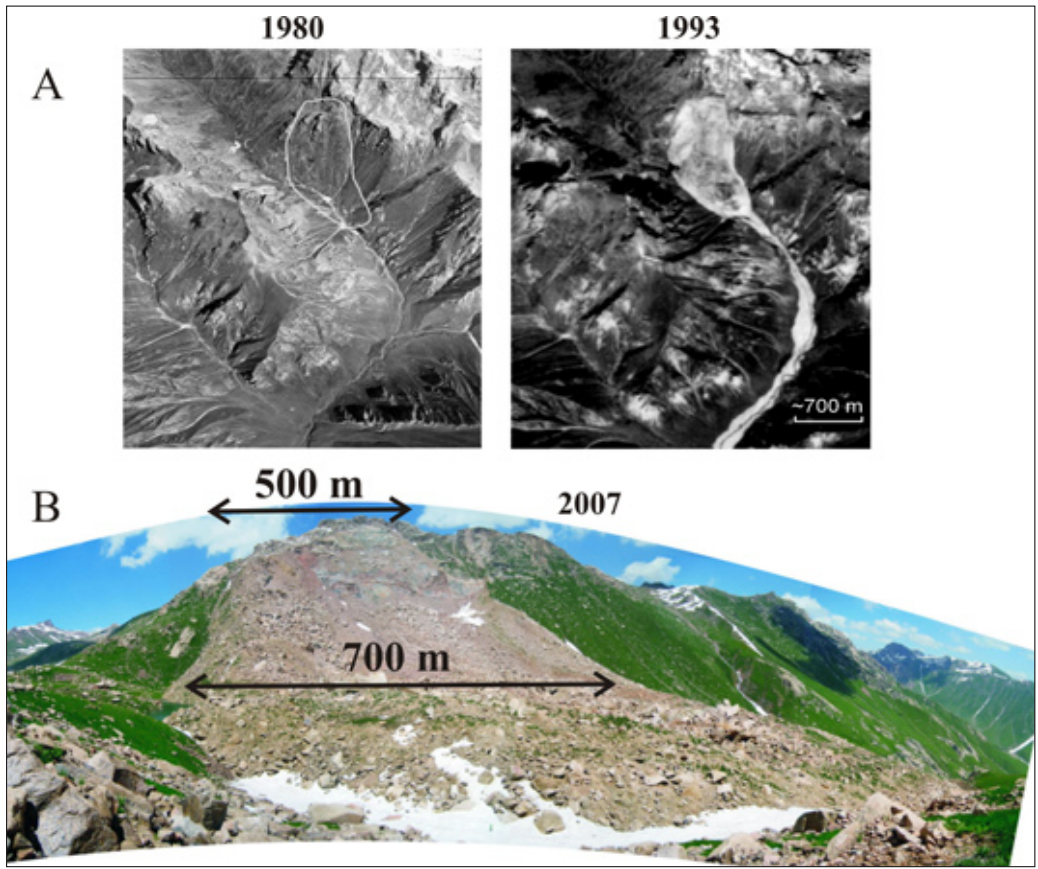

**Figure 8:** The Belaldy rock avalanche. A) Aerial photograph of the site acquired in 1980 (left), before the 1992 Suusamyr earthquake and space image of 1993 (right), after the earthquake. B) Photograph of the rock avalanche (scarp, behind, and dam, in front) in 2007.

*3.1.6. The Kainama earth-flow – a delayed effect of small regional earthquakes?*

The Kainama earth-flow (220 000 m$^3$) developed on April 26, 2004, had caused the destruction of 12 houses and the death of 33 people (64 families became homeless) south of the Fergana Basin (location in Fig. 1). This landslide had formed within the loess layer and developed into a very rapid flow with a long runout. Fig. 9 shows that, in 2001, no clear sign of instability could be seen. In spring 2003, several cracks appeared on the upper part of the slope.

The main failure of 2004 April was preceded by an increased seismic activity: on 2004 March 26, an earthquake of Ms=4.7 occurred 40 km SW of the Kainama site (PGA at the site ~ 0.05 g). Two other smaller earthquakes (Ms=4.2) were registered some 50 km NW of Kainama on April 8 and 9. New fractures had been observed after these earthquakes.

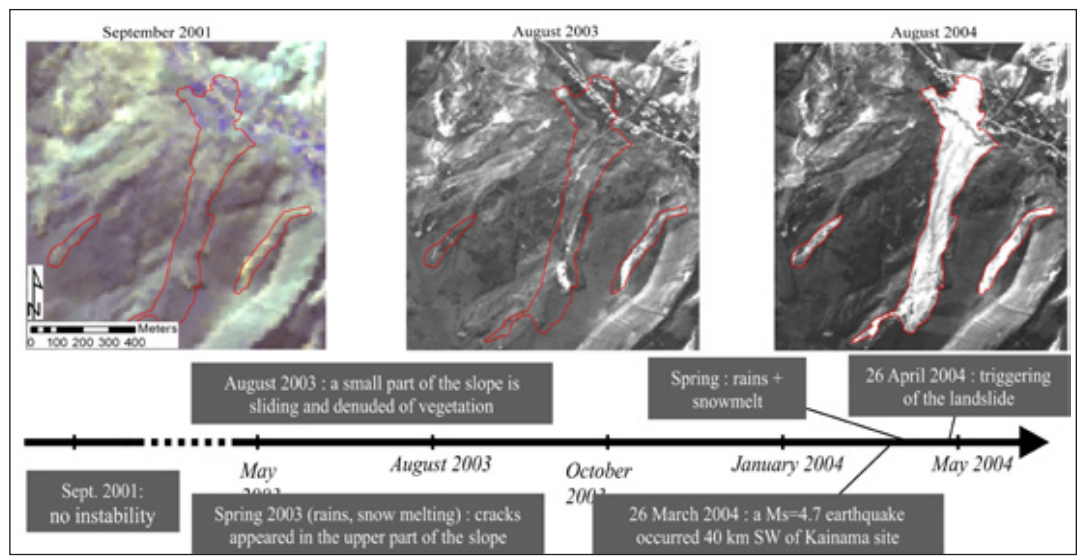

**Figure 9:** Evolution of the Kainama landslide from 2001 (ASTER satellite image, left), to 2003 (SPOT image, middle) and 2004 (SPOT image, right).

The delay of massive slope failure after the earthquakes is likely to be related to changes of the local environmental conditions. At the end of March the mountains were still covered by snow – and the groundwater level was low. Snowmelt and spring rains in April caused a rapid increase of groundwater level – probably enhanced by rapid water infiltration through the new fractures induced by the earthquakes.

To confirm the important role of this process-sequence for development of many landslides in seismic areas, numerical modelling studies are being undertaken. Preliminary results had been published in Danneels et al. (2008).

### 3.2. The 1920 Haiyuan (China) and 1970 Peru events

As shown previously, earth-flows in loess deposits and rock-debris avalanches proved to be the most catastrophic mass movements in Central Asian mountain regions. Therefore, the 1920 Haiyuan and 1970 Peru earthquakes were selected for comparison since they are known to have triggered, respectively, the most disastrous loess earth-flows and the most catastrophic single rock-ice avalanche.

#### 3.2.1. The Haiyuan or Gansu earthquake, 1920

On December 16, 1920, a M=8.5 earthquake occurred near Ganyan Chi, Haiyuan County of the Ningxia Hui Autonomous Region in China (Zhang, 1995). Several hundreds of thousands of houses collapsed and officially 234117 people died. Zhang (1995) noticed that particularly high intensities were recorded over areas covered by thick loess deposits: '…areas with intensity of 8 … here the damage caused by the slides was more serious than the primary ones caused by the quake. Based on field survey, abnormal 9 to 10 intensities were recorded here'. Zhang (1995) added that 'these landslides were not only controlled by the intensity of the earthquake, but by the structure of the subsoil'. It should be noticed that such an intensity of 8 (on Chinese Seismic Intensity Scale, CSIS – comparable to the European Macroseismic Scale, EMS-98) was recorded over 50000 km$^2$ (Zhang, 1995). For comparison Zhang (1995) noted also that '…in spite of the high intensity of the quake, individual slides were quite small, their density low in this mountainous area, thanks to the thin soil cover and the bedrock is often exposed.' In total, a surface area of more than 4000 km$^2$ was affected by landslides, which buried numerous villages.

Zhang & Wang (2007) referring to Close and McCormick (1922), reported that about 100000 people were killed by landslides in loess deposits. Dangjiacha landslide (Fig. 10) was one of the most catastrophic mass movements triggered by the 1920 earthquake. The landslide formed a dam with a volume of about 15 10$^6$ m$^3$. Behind the dam, the largest lake induced by the earthquake had been impounded.

Zhang & Wang (2007) observed that loess earth-flows triggered by the Haiyuan earthquake had developed on relatively gentle slopes compared to those triggered by rainfall in the same region.

These observations highlight the particular susceptibility of loess areas to ground failure, such as it was clearly shown by Derbyshire et al. (2000) analysing geological hazards affecting the loess plateau of China. The failure mechanisms of loess earth-flows and their connection with liquefaction phenomena will be analysed more in detail in the next section.

#### 3.2.2. The Peru earthquake, 1970

On May 31, 1970, the M=7.7 Peru earthquake triggered one of the most catastrophic landslides that have ever occurred: the Nevado Huascaran rock-ice avalanche. This event is not only known for the large number of casualties it has caused, about 18000, but also for its geologically fascinating aspects (Plafker et al., 1971). The huge mass of rock and ice with a volume of more than 50 10$^6$ m$^3$ originated from the west face of the 6654 m high north peak of Nevado Huascaran at some 130 km east of the earthquake epicentre. The particularity of this debris avalanche is its long travel distance of about 16 km.

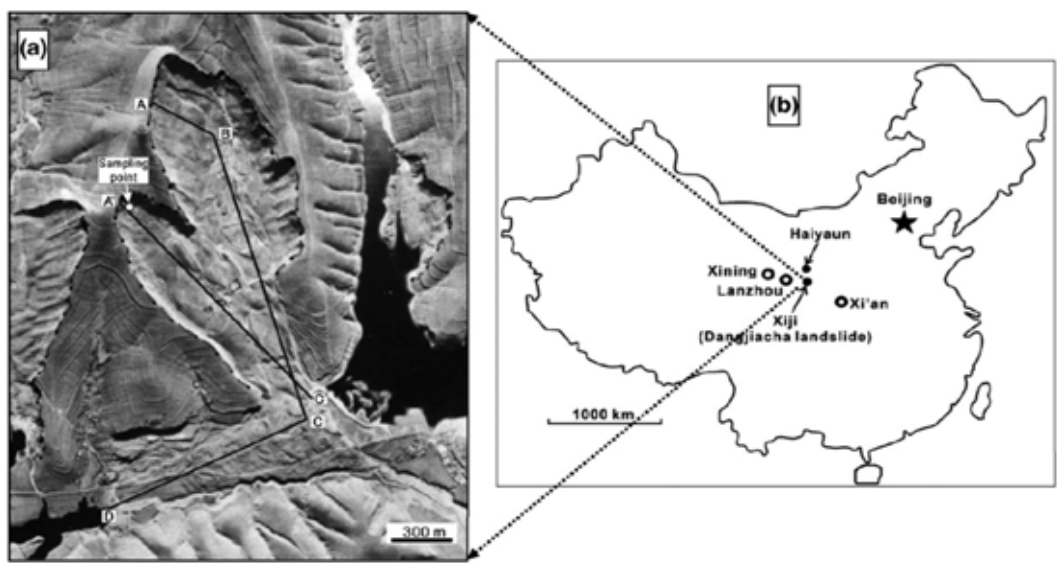

**Figure 10:** Dangjiacha landslide (a) and its location (b), after Zhang & Wang (2007).

Before reaching the villages of Yungay and Ranrahirca, the mass slid for more than 2 km over a glacier, filled the valley floor at the foot of the slope and then was channelled through a gorge. Just before the end of the gorge, the avalanche split into two lobes one moving on towards the village of Ranrahirca, one overtopping the 230 m high southern ridge of the gorge and continuing its way to Yungay village. Peak velocities of this rock avalanche were actually estimated at more than 100 km/h by Erismann & Abele (2001) and even to about 300 km/h by Plafker et al. (1971).

### 3.3. Preliminary conclusions based on the case histories

The Nevado Huascaran and the Khait mass movements show that volume, runout and speed are three important factors contributing to the impact potential of mass movements. The volume and runout of the Khait rock avalanche (40 $10^6$ m$^3$, about 9 km) are somewhat smaller than for Nevado Huascaran rock-debris avalanche (>50 $10^6$ m$^3$, and 16 km). Further, no ice and probably no (free) water were involved in the Khait rock avalanche, such as in the Nevado Huascaran mass movement favouring its fluidisation. However, according to Evans et al. (2007), significant masses of loess were entrained along the path of the Khait rock avalanche. The mobility of this loess outlined above is likely to have contributed to the speed of the mass movement. We believe that the Khait rock avalanche was travelling with a speed similar to the one of the Nevado Huascaran when it reached the valley. The high speed of the Khait rock avalanche is also mentioned in the following notes from Solonenko (1977): '…Its rate was about 100 km/h… A powerful airwave went before the collapse. It broke constructions, uprooted trees and threw them down for hundreds of metres.'

The Khait rock avalanche is one of the largest (recent) runout mass movements in the Tien Shan. Generally, rockslides with a similar volume (e.g. Belaldy rock avalanche) hit the opposite valley flank and stop or move only a few hundreds of meters along the valley floor where they form a landslide dam. A debris flow developing from a dam breach may, however, have a very long runout – about 20 km in the case of the partial Belaldy rockslide dam failure.

Similarly, loess earth-flows are known to have caused rapid and long runout mass movements. For instance, in the Dzhalal-Abad region, Fergana Valley, Kyrgyz Republic, loess earth-flows had travelled over 7 km in 1994 killing several tens of people.

Considering also the previous notes on geological disasters in loess deposits, we believe that earthquake and landslide hazards are particularly high in those regions of Central Asia, which are covered by several tens of meters of loess: the foothill regions of the Fergana Basin rim (Kyrgyz Republic and Tajikistan) and around the town of Dushanbe (Tajikistan) as well as mountain valleys covered by loess up to an altitude of 2000-2500 m.

However, a disaster such as the 1920 Haiyuan earthquake is not expected to hit any of those regions for two reasons: first, a M>7.5 earthquake is very unlikely to occur in these areas; second, the loess cover is significantly thinner than on the loess plateau of China where the thickness can reach 300 m (Derbyshire et al., 2000).

Considering the enormous impact potential of massive rockslides, such as Belaldy or Khait, it is important to assess their occurrence probability. This probability is closely related to the recurrence time of the triggering earthquake (actually, only very few massive non-seismic rock avalanches are known from the Tien Shan) – but also other, climatic, aspects have to be considered as shown below.

First, for the Suusamyr region, it can be estimated according to the earthquake statistics presented by Abdrakhmatov et al. (2003) that the 1992 earthquake has a 500-year return period. Therefore, in near future, a similar earthquake – and a related rock avalanche such as Belaldy - is not expected to hit the same region. However, for the entire Tien Shan and Northern Pamir region, the analysis of Abdrakhmatov et al. (2003) showed that a return period of less than 20 years can be estimated for a M≥7 earthquake – able to trigger massive rockslides. Considering that the last M≥7 earthquake occurred in 1992, a rough computation of conditional probability indicates that there is a 90% chance to exceed a 7-magnitude event within the next 10 years. The probability to have such an earthquake – and related mass movements - seems to be highest (with a chance of 90% to exceed a M=7 earthquake within the next 10 years) along the Tien Shan – Pamir boundary region, where the last M≥7 earthquake was the Markansu event in 1974 close to the Tajik-Kyrgyz-Chinese border.

Recently on October 5, 2008, a M=6.6 earthquake hit this 3-border region and killed 75 people in Nura village, Kyrgyz Republic, which was entirely destroyed. A quick analysis of a SPOT 2.5 m resolution image acquired on October 16, 2008, eleven days after the earthquake showed that this earthquake had not triggered any major mass movements, besides a few rock falls in remote high mountain areas. We believe that the time of the earthquake after the dry summer season favoured slope stability. Also, we estimate that the Suusamyr earthquake in 1992 had triggered relatively few and small landslides due to its occurrence in August at the end of the dry summer season.

Though the 2008 Nura earthquake had been a major event, we think that the M≥7 earthquake along the Kyrgyz-Tajik border is still to come and most likely before 2020. According to previous observations, such an earthquake is most likely to trigger many large mass movements in early summer season (May-July), after snowmelt and spring rains.

## 4. Influence of seismic shaking and long-term effects

In the following, we will outline a series of dynamic processes involved in seismic ground and slope failure and long-term mass movement hazards.

## 4.1. Shaking effects

### 4.1.1. Shaking intensity, orientation and amplification

First, we will review some general observations of the influence of shaking on slope instability. The fact that earthquake-triggered landslides occur preferentially along activated faults and within the hanging wall is closely connected to the stronger shaking intensity in these parts close to the faults. Keefer (1999) reviewed shaking thresholds for the three main mass movement classes. Using the Arias Intensity (computed on the basis of the integral of the square of acceleration of the recorded signal, see Arias 1970) as shaking unit, he showed that disrupted slides would have the lowest threshold of about 0.1 – 0.15 m/s, coherent slides would be triggered for a shaking of more than 0.3-0.5 m/s and lateral spreads and flows for more than 0.5 m/s.

Second, we will analyse the influence of certain shaking particularities on slope stability. Considering the displacement of mass movements during the Wenchuan earthquake, Yin et al. (2009) concluded that for some types, which occurred very close to the fault, the vertical shaking component had significantly contributed to the failure. Actually, they observed that close to the fault ground motion records indicated a higher vertical than horizontal acceleration. It should be noticed that there are only few of such observations and that more research is needed to assess the effect of vertical shaking on slope failure. Another influence of shaking orientation has been pointed out by Wen et al. (2004) referring to Li (1978) who noticed that the largest landslides occur when ground shaking and slope aspect are in the same direction. To better quantify these observations, it would be necessary to investigate them by numerical simulations.

A very common observation already highlighted above is related to the topographic effects already mentioned above: seismic landslides are preferentially triggered from convex surface morphologies and hillcrests, which are also particularly prone to fracturing during earthquakes.

To better assess this effect at the Las Collinas landslide site, Crosta et al. (2004) performed ambient noise measurements. Comparing the horizontal to vertical ratios on the Balsamo ridge from where the landslide was triggered with those at the bottom of the ridge, they observed an amplification of about 1.4 at the ridge.

We have extensively investigated this 'topographic' effect on seismic failure triggering (e.g. Havenith et al., 2003a; Bourdeau & Havenith, 2008) and observed that, generally, the morphological contribution to the amplification on ridge-tops is much smaller than the effect of the local geology. For rock slopes, Havenith et al. (2002) showed that stronger and deeper weathering below convex surface morphologies reducing the elastic shear modulus of the material is the main factor favouring seismic ground motion amplification. For soft sediment sites, it was observed that a thicker cover of loess on top of slopes could also induce strong amplification (Danneels et al. 2008). We, therefore, suggest using the term 'site effects' well-known from the earthquake engineering approach to describe seismic amplification at the surface – site effects describe both wave focusing effects due to the morphology ('topographic effect') and due to the local geology (often called 'surface layer amplification').

### 4.1.2. Liquefaction and processes in rapid loess earth-flows

Jurko & Sassa (2008) proposed to use the term liquefaction only to describe a rapid loss of strength resulting from increased pore water pressure and reduced effective normal stress.

Ishihara (2002) reanalysed the landslide event triggered in Gissar, Tajikistan, by a M=5.5 earthquake in 1989. He concluded that extensive liquefaction must have developed in the loess deposit in the gently sloping hilly terrain to produce such catastrophic earth-flows. Zhang & Wang (2007) related the high mobility – strongly contributing to the overall impact of the mass movements - of the loess earth-flows triggered by the Haiyuan earthquake in 1920 to liquefaction phenomena. Through geotechnical tests, they also proved that the reduction of shear resistance mainly results from the generation of pore water pressure.

Until the Gissar event, liquefaction was known to affect water-saturated sand deposits (due to the larger grain size of sand with respect to loess). And Ishihara (2002) considered that the liquefaction in Gissar was unique and had occurred unexpectedly in a windlaid deposit of silt in a semi-arid region. Now, we know that exactly the same mechanism had mobilised hundreds if not thousands of landslides during the Haiyuan earthquake in the Gansu Province, China. Actually, Jurko & Sassa (2008) showed that not only sandy but also weakly plastic silt containing some clay could exhibit liquefaction under saturated undrained cyclic loading conditions.

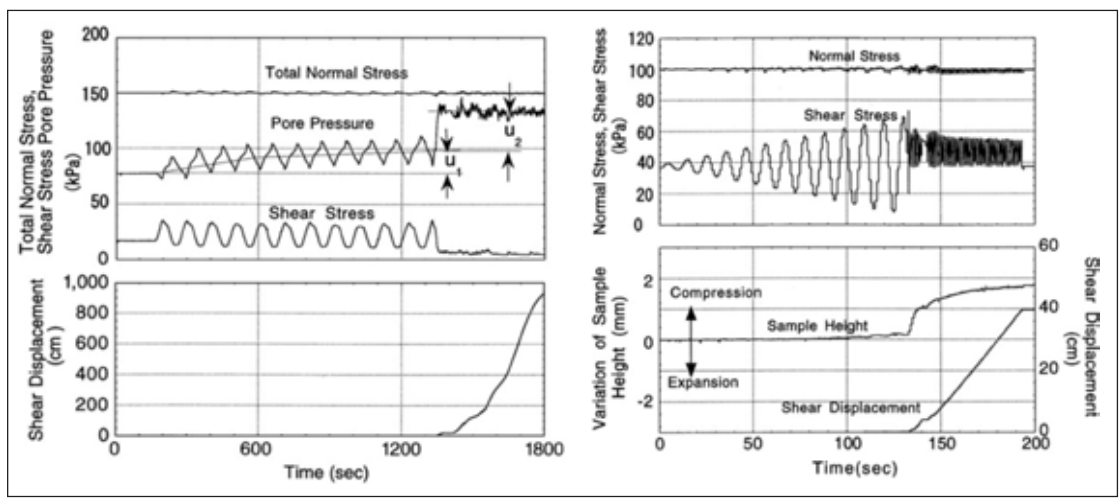

**Figure 11**: Cyclic loading tests applied to saturated (left) and dry (right) loess samples (after Zhang & Wang, 2007).

Ishihara (2002) attributed the liquefaction of the loess deposits in Gissar to their collapsible nature and the high water saturation at depth, related to infiltration of irrigation water over many years. The low plasticity of the loess contributes to its liquefaction under seismic shaking and its development into an earth flow with runouts of several kilometres. Recently, Jurko & Sassa (2008) suggested using the plasticity index (difference between liquid and plastic limit) as a criterion for estimating the liquefaction potential of clayey soil.

Investigating the phenomenon of liquefaction in clayey soils, Osipov et al. (2005) showed that: 1) low plasticity kaolin-, illite- and bentonite-sand mixtures were very vulnerable to liquefaction; 2) an increase in bentonite content raised both the soil plasticity and its resistance to liquefaction; and 3) the bentonite-sand mixtures with a plasticity index larger than 15 (i.e. requires a large amount of water to turn from a plastic to a liquid behaviour) were resistant to liquefaction.

Wang et al. (2007) studied the Tsukidate landslide in sandy soils triggered by an earthquake in 2003 in Japan. They showed that under undrained conditions seismic loading results in a certain amount of excess pore water pressure within the saturated sliding surface, and then leads to the failure of the slope. After failure, high excess pore water pressure was generated with increase of shear displacement – a phenomenon called 'sliding surface liquefaction' (Sassa et al., 2005). This finally resulted in a significant reduction of the shear resistance and in rapid movement.

Zhang & Wang (2007) showed that, during a cyclic ring shear test on a saturated loess specimen under undrained conditions, cyclic shear stress clearly induced increasing pore pressure (u1 in top left part of Fig. 11). At the moment of failure, a quick generation of high pore pressure was observed (u2 in Fig. 11). And the cyclic ring shear test on loess with natural water content confirmed that this quick generation resulted from the instant collapse of soil structure at failure. After failure, the steady state was reached, where shear displacement continued under constant effective stress and shear resistance (right part in Fig. 11). Tests with dry loess samples showed that shear displacement increased only during cyclic loading, no additional displacement was generated after the loading. Zhang & Wang (2007) concluded that in a slope consisting of dry loess, displacement could be generated only during the earthquake, and no long runout landslide would occur.

### 4.2. Long-term effects

The traces of strong earthquakes in the form of fault scarps and mass movements may be well preserved in the local relief and even be used for paleo-seismic studies. Korjenkov et al (2002) showed that some morphological effects of earthquakes in the Chilik-Kemin fault zone could be dated back to Holocene, Late Pleistocene and even Late-Middle Pleistocene through geological and geomorphologic correlations with the local, relative Pleistocene chronology of river terraces and glacial moraines. For the Suusamyr region, Korjenkov et al. (2004) could outline that several newly triggered landslides and gravitation cracks had formed within paleo-seismic mass movements. On the southern slope of Chet-Korumdy ridge he could define three different generations of surface deformations – the third one being related to the Suusamyr earthquake, while the two others are remnants of old events, most likely similar strong earthquakes.

Nepop & Agatova (2008) investigated large mass movements in the Altai Mountains. They found that the large mass movements leave the most persistent imprint on landforms and thus represent the longest period of seismic activity.

These studies show that earthquakes may have long-lasting effects on the landscape, which can be used for paleo-seismological studies. Consequently, they also do have mid-term effects – some of which may have significant impact on the global hazard.

Some mid- and long-term effects have clearly been outlined for the 1999 Chi-Chi and 2005 Kashmir earthquakes. Dadson et al. (2004) analysed the co- and post-seismic geomorphic impact of the 1999 Chi-Chi event in Taiwan. In addition to the 20000 landslides immediately triggered by the earthquake, they found that co-seismic weakening of substrate material had caused increased landsliding during subsequent typhoons. Further, they observed that most of the co-seismically produced landslide material was transported towards the rivers, and thus increased sediment concentration during the storms after the earthquake.

Petley et al. (2006) observed that the Kashmir earthquake had produced a 'very large extent of slope cracking in areas within 5 km of the fault'. These cracks would represent the initial failure type of future landslides, which had not developed into massive failure due to the low groundwater level. And, indeed, the precipitation level during the months before the earthquake had been anomalously low. Further, they related the more intensive landslide activity during the 2006 snowmelt and monsoon seasons to the 'extensive fissuring produced during the earthquake'.

Keefer (1999) also analysed the long-term effects of earthquakes outlining that large earthquakes may modify the characteristics of drainage basins. Erosion and transport of the landslide material may then finally result in creation of new alluvial fans, increasing also the potential of debris flows or water flows over time.

## 5. Conclusions

A series of case histories of earthquake-induced landslides in Central Asia have been presented. The review clearly pointed out that, in these regions, large rockslides and rapid earth-flows in loess deposits have the highest impact potential. Therefore, some comparisons have been made with worldwide known earthquakes events, which had triggered disastrous slope failures of similar type, in particular the 1920 Haiyuan earthquake in China and the 1970 Peru earthquake. The volume and the mobility (runout and speed) of the mass movements play an important role for related hazard and risk. In this regard, it is not surprising that the most disastrous mass movement known from Central Asia is a long runout rock avalanche of a volume of about $40 \cdot 10^6$ m$^3$, the Khait rock avalanche, which is supposed to have reached a speed of more than 100 km/h. Note, several massive and rapid rockslides occurred also in remote areas of Central Asian regions, where nobody or only few persons had been killed (e.g. during the Kemin earthquake, 1911).

While most of these giant rockslides have been triggered by large magnitude seismic events (M ≥ 7) in Central Asian mountain regions, loess earth-flows may also be triggered by smaller earthquakes – or even by climatic factors alone. Here, we presented some examples of fatal loess landslides triggered directly by a M=5.5 earthquake in Tajikistan and one landslide, which had developed in loess deposits several weeks after a M=4.7 earthquake. The comparison with the Haiyuan earthquake event of 1920 showed that such loess landslides could be very disastrous.

For loess earth-flows, the high impact potential is, however, not only related to the volume and mobility of a single landslide - which are generally lower than for massive rock avalanches. Here, the high spatial and temporal occurrence probabilities clearly contribute to related risk. The 1989 Gissar earthquake case history has shown that also M=5.5 earthquakes are able to trigger numerous loess landslides. Such earthquakes occur almost every year in the Central Asian mountain regions - many of which are covered by several meters of loess, especially in the foothill areas. For comparison, it was shown that the return period of a M ≥ 7 earthquake in the Tien Shan and northern Pamir able to trigger massive rockslides is about 20 years.

The importance of mid- and long-term effects was outlined both for rockslides and loess landslides. Several case histories showed that one important – if not the most important – long-term consequence of massive rockslides can be the formation of a dam and the impoundment of a natural reservoir. Actually, the largest still existing rockslide dam on earth had formed in 1911 in the Pamir Mountains. The worst-case scenario flood wave triggered by dam failure could affect more than 5 million people living in the Amu Darya river basin. The 1992-1993 Belaldy rockslide case history also documents the risk due to dam formation; partial failure of the dam resulted in a 20 km long mud- and debris-flow, which caused significant damage downstream.

For loess landslides, it was shown that seismic ground motions must not necessarily be the final trigger, but could also be a preparatory factor of slope failure. Relatively weak seismic shaking (e.g. < 0.1 g) is able to produce ground fractures, but often without inducing a mass movement. However, these ground fractures can facilitate water infiltration and related increase of groundwater pressures, which could lead to slope instability.

In this regard, it was also shown that similar earthquakes may not necessarily trigger the same amount of landslides, due to different climatic conditions and groundwater level at the time of the earthquake: the M=6.6 Nura earthquake in 2008 triggered relatively few landslides due to the occurrence at the end of the dry summer season. However, we do not know yet if snowmelt and spring rains in 2009 have triggered any new landslides.

Finally, the case histories showed that related landslides were not only instantaneous effects of earthquakes – some had already developed before the seismic shock and some continued or started moving well after the shaking. To better assess the short- to long-term effects earthquakes on slopes, landslides need to be monitored by geophysical, seismological and geotechnical systems, coupled to multi-temporal satellite imagery and numerical modeling of multi-event scenarios. In the frame of new projects on landslide problems in Central Asia, focus will be on the installation of such monitoring – modeling systems.